\documentclass[prd,twocolumn]{revtex4}%
\usepackage{amssymb}
\usepackage{amsmath}
\usepackage{amsfonts}
\usepackage{graphicx}%
\begin{document}
\title{Long-distance behavior of the quark-antiquark static potential. Application to
light-quark mesons and heavy quarkonia.}
\author{P. Gonz\'{a}lez}
\affiliation{Departamento de F\'{\i}sica Te\'{o}rica, Universidad de Valencia (UV) and IFIC
(UV-CSIC), Valencia, Spain.}
\date{\today}

\begin{abstract}
Screening effects from sea pairs on the quark-antiquark static potential are
analyzed phenomenologically from the light-quark to the heavy-quark meson
spectra. From the high excited light-quark meson spectrum a universal form for
the screened static potential is proposed. This potential is then succesfully
applied to heavy quarkonia. Our results suggest the assignment of $X(4260)$ to
the $4s$ state of charmonium and the possible existence of a $5s$ bottomonium
resonance around 10748 MeV.

\end{abstract}

\pacs{14.40.-n,12.39.-x}
\maketitle

\section{Introduction}

One remaining problem in our understanding of QCD has to do with quark
confinement in hadrons. We expect confinement to be the dominant
quark-antiquark ($q-\overline{q})$ or quark-quark interaction at large
separation distances and therefore to be determinant to explain the properties
of highly excited (large sized) hadrons. In recent years there has been an
important progress in the knowledge of the spectrum of highly excited mesons
in the light-quark ($u,d)$ \cite{Bug04} as well as in the heavy-quark ($c,b)$
\cite{PDG08,Swa06} sectors. In particular highly excited light-quark mesons
show an intriguing hydrogen-like spectral pattern \cite{Afo07} which can be
reproduced, within a non relativistic constituent quark model framework, by
means of a static $q-\overline{q}$ interaction which becomes asymptotically
coulombic \cite{Gon08}. Here we suggest that this asymptotic behavior has to
do with confinement (non-perturbative gluonic effects) and not with the
perturbative gluonic coulomb interaction as suggested in \cite{Gon08}. We
propose that string breaking gives rise, from a linear confining interaction,
to an asymptotically coulombic potential. When this screened confinement is
complemented with a screened \textquotedblleft gluonic\textquotedblright%
\ coulomb interaction an accurate description of the highly excited
light-quark meson spectrum is achieved. The resulting static potential depends
on four parameters. The two of them entering in the confinement term, the
string tension and the string breaking distance,\ are considered as universal
in the sense of having the same values in all meson sectors. This is also the
case for the orbital-angular-momentum parameter related to the onset for
states of confinement. The remaining parameter, the effective
\textquotedblleft gluonic\textquotedblright\ coulomb strength, gets different
values when going from light to heavier quarks. It should be pointed out that
the potential does not contain any additive constant. The calculated meson
masses, obtained by adding the mass of the quark and the mass of the antiquark
to the eigenvalues of the Schr\"{o}dinger equation, are directly compared to
the experimental meson masses. In practice, for equal quark and antiquark
masses, the quark mass and the effective \textquotedblleft
gluonic\textquotedblright\ coulomb strength are fixed from two well
established experimental meson masses in the region of applicability of the
static approach.

The successful spectral description obtained in the light-quark meson case can
be extended to other meson sectors, in particular to heavy quarkonia where an
accurate description of the highly excited states may be very helpful for an
unambiguous quantum numbers assignment. Our results suggest that $X(4260)$
could be the $4s$ state of charmonium and that a non-cataloged bottomonium
resonance around $10748$ MeV might exist.

These contents are organized as follows. In Section II we establish the
general character of the constituent quark model approximation we use and the
general criterion of validity of the static potential in the different meson
sectors. In Section III the asymptotically coulombic potential inferred from
the study of the highly excited light-quark meson spectrum is derived from a
screened confinement potential ansatz. The implementation of an effective
\textquotedblleft gluonic\textquotedblright\ interaction and the consideration
of an additional $L-$ dependent correcting factor for it allows for a precise
description of the known static spectrum. In Section IV the same potential is
applied to heavy quarkonia, charmonium and bottomonium, where a distinctive
quantum numbers assignment for high excited states comes out.\ Finally in
Section V we summarize our main results and conclusions.

\section{Quark model approach}

In the study of the meson spectra, from light to heavy quarkonia, we shall
rely on a Non Relativistic Constituent Quark Model (NRCQM) framework. We will
solve the Schr\"{o}dinger equation for a static potential. Although the
application of the NRCQM\ to heavy-quark systems, at least for bottomonium,
can be taken for granted, its application to light-quark systems ($m_{u}=340$
MeV) is always a matter of debate. In the spirit of NRCQM calculations the
effective values of the parameters take into account, at least to some extent,
relativistic corrections in the kinetic and potential energies. Actually it
has been recently shown \cite{Gon08} that the known spectrum of highly excited
light-quark mesons can be nicely reproduced within such a calculation
framework despite its very relativistic character indicated by the calculated
values of $p_{u}/m_{u}\geq1$.

The main distinctive feature of the effective potential employed in
\cite{Gon08} is its asymptotically coulombic tendency. Explicitly%
\begin{equation}
V_{light-quark}(r\rightarrow\infty)=\sigma_{u}r_{s}-\frac{k_{u}}{r}+C_{u}
\label{ec1}%
\end{equation}
where $\sigma_{u}=932.7$ MeV/fm stands for the string tension, $k_{u}=2480$
MeV.fm for a coulomb strength and $C_{u}=1070$ MeV for a constant to fix the
origin of the potential. The distance $r_{s}$ represents the onset for the
screening of the interaction due to the presence of light quark-antiquark
pairs popping out of the vacuum. The value chosen $r_{s}=1.15$ fm is inferred
from lattice calculations (see \cite{Bal01} and references therein). As for
$\sigma_{u}$ the value used is extracted from the phenomenological analysis of
the ($\rho,$ $a_{2},...)$ Regge trajectory, see also \cite{Bal01}. Concerning
the value of $k_{u}$ one could tentatively try to ascribe it to the
chromoelectric one gluon exchange (OGE) interaction as done in \cite{Gon08}.
However, as we shall show in the next section it may be rather giving account
of the long distance attenuation, due to string breaking, of the linear
confining term.

It should be reminded that the accurate energy description of the meson states
is linked to the correct long distance behavior of their wave functions. Given
the relativistic character of the fitted spectrum we should not trust much, at
intermediate and short distances, the non-relativistic wave functions obtained
from the Schr\"{o}dinger equation. Only very large sized light-quark mesons
(in our scheme the higher the root mean square (rms) radius of the meson the
lower the $p_{q}/m_{q}$ value), for which there are no available data yet, can
be considered non-relativistic systems. For them the wave function coming out
from the Schr\"{o}dinger equation might also be accurate at short and
intermediate distances.

For the sake of completeness it is worthwhile to remind the criterion derived
in \cite{Gon08} for the applicability of a static potential to a meson
($q\overline{q})$ sector within our NRCQM framework. It reads%
\begin{equation}
<r^{2}>^{1/2}>>\frac{1}{m_{q}} \label{ec2}%
\end{equation}
where $<r^{2}>^{1/2}$ stands for the root mean square (rms) radius of the
meson state. So only for mesons with a large size, as compared to $1/m_{q}$,
the static approach makes sense. For $u$ and $d$ quarks ($m_{q}=340$ MeV) this
means $<r^{2}>^{1/2}>>0.6$ fm. In fact the light-quark meson spectrum has been
well reproduced for states with rms radii greater or equal than three and a
half times this limit: $<r^{2}>^{1/2}\gtrsim2.1$ fm.

When going to heavier quarkonia we have $<r^{2}>^{1/2}>>0.4$ fm for
$s\overline{s}$ ($m_{s}\sim500$ MeV), $<r^{2}>^{1/2}>>0.14$ fm for
$c\overline{c}$ ($m_{c}\sim1400$ MeV) and $<r^{2}>^{1/2}>>0.04$ fm for
$b\overline{b}$ ($m_{b}\sim4800$ MeV) where typical values for the constituent
quark masses have been chosen. Then, by using the same validity factor of 3.5
as in the light-quark case, we expect the static approximation to be valid for
$(<r^{2}>^{1/2})_{s\overline{s}}\gtrsim1.6$ fm, $(<r^{2}>^{1/2})_{c\overline
{c}}\gtrsim0.6$ fm and $(<r^{2}>^{1/2})_{b\overline{b}}\gtrsim0.16$ fm.

Notice though that the constituent quark mass in a meson sector is in our
model a parameter to be fixed from data and that the values of $<r^{2}>^{1/2}$
result from the solution of the Schr\"{o}dinger equation. Therefore the
established criterion has to be checked \textit{a posteriori}. Nonetheless its
consideration is essential to adequately select the specific data to be used
to fix the free parameters. Thus, for light-quark mesons, data corresponding
to $L=4$ and $L=5$ states were used since these states are expected to have
large rms radii due to the presence of the centrifugal barrier.

\section{Light-quark mesons}

\subsection{String breaking}

The static $q-\overline{q}$ potential can be derived from Lattice QCD
\cite{Bal01}. In the quenched approximation, only valence quark $q_{v}$ and
antiquark $\overline{q}_{v},$ it has the funnel form%
\begin{equation}
\overline{V}(r)=\sigma r-\frac{\zeta}{r} \label{ec3}%
\end{equation}
where $\sigma$ is the string tension and $\zeta$ is the strength of the
coulomb interaction. This potential has to be corrected at short distances so
that $\zeta$ becomes a function of $r$ \cite{Kuz02}$.$ When including sea
quarks, an unquenched potential results from the screening of the static
sources $q_{v}$ and $\overline{q}_{v}$ by light $q\overline{q}$ pairs created
in the hadronic vacuum. A parametrization of this effect was proposed twenty
years ago \cite{Bor89}. Unquenched lattice results for the potential between
two heavy static quarks separated by a distance $r:0\rightarrow1$ fm were
described by the potential%
\begin{equation}
\overline{V}_{scr}(r)=\left(  \sigma r-\frac{\zeta}{r}\right)  \left(
\frac{1-e^{-\mu r}}{\mu r}\right)  \label{ec4}%
\end{equation}
where $\mu^{-1}$ represented a screening length and $\zeta$ was related to the
quark-quark-gluon coupling $\alpha_{s}$ through $\zeta=(4/3)\alpha_{s}$. The
screening factor $H(r)\equiv\left(  \frac{1-e^{-\mu r}}{\mu r}\right)  $ was
constructed so that $\overline{V}_{scr}(r)$ has a coulombic behavior at small
distances whilst approaching a constant at large distances. When applied to
heavy quarkonia this potential form with effective values of its parameters
provided a precise description of the spectrum of $b\overline{b}$ states with
rms radii smaller than $1.1$ fm \cite{Din95,Gon03}. However, up to now,
lattice calculations do not allow to extract the precise form of the QCD
static potential at large distances \cite{Bal01,Dun01}. So the asymptotic
constant behavior should be considered as an educated guessing.

Alternatively an attenuated linear form of confinement has been implemented
for the asymptotic potential in the framework of the QCD string approach
(QCDSA) that has been successfully applied to light-quark \cite{Bad02,Sim02}
and heavy-quark \cite{ccBad08,bbBad09} mesons. In this physical picture the
string tension $\sigma$ is attenuated at separations $r\gtrsim R_{1}\simeq1.2$
fm becoming a function of $r$ so that for $r\gtrsim R_{2}=2.5$ fm string
breaking occurs with large probability. This attenuation plays a key role to
correctly obtain the masses of the radial excitations of light-quark mesons
within this approach. More explicitly the confinement potential reads
\begin{equation}
V_{SA}(r)=\sigma(r)r=\sigma r\left(  1-\gamma\frac{\exp\left(  \sqrt{\sigma
}(r-R_{1}\right)  }{B+\exp\left(  \sqrt{\sigma}(r-R_{1}\right)  }\right)
\label{ec5}%
\end{equation}
with $\sigma=0.185$ GeV$^{2}=937.5$ MeV/fm, $\gamma=0.4,$ $R_{1}=6$
GeV$^{-1}=1.18$ fm and $B=20.$ The screening factor between parenthesis that
will be called $G(r)$ henceforth varies from $\simeq1$ for $r=0$ to a value of
$\simeq(1-\gamma)$ for $r>R_{2}.$

Following the same philosophy as in \cite{Sim02} we shall attempt to extract
information over the variation of the confining potential with $r$ from a
systematic analysis of the meson spectrum within our NRCQM framework. As
mentioned above the main distinctive feature resulting from the application of
the NRCQM to the light-quark meson spectrum is the coulombic asymptotic
behavior of the potential as given by Eq.(\ref{ec1}). It is then interesting
to examine the possibility that it may come from confinement as a result of
string breaking. Indeed the form of the potential in Eq.(\ref{ec1}) can be
derived from the screened confinement potential ansatz%
\begin{equation}
V_{conf}(r)=\sigma r(1-e^{-\frac{\nu}{r}})\equiv\sigma rF(r) \label{ec6}%
\end{equation}
as can be easily checked by using $F(r)\rightarrow\frac{\nu}{r}-\frac{\nu^{2}%
}{2r^{2}}$ and making the identifications (up to order $1/r^{2}$)%
\begin{equation}
\sigma\nu=\sigma_{u}r_{s}+C_{u} \label{ec7}%
\end{equation}%
\begin{equation}
\frac{\sigma\nu^{2}}{2}=k_{u} \label{ec8}%
\end{equation}
Then from the numerical values of $\sigma_{u},$, $C_{u}$ and $k_{u}$
previously quoted we get%
\begin{equation}
\nu=2.3\text{ fm} \label{ec9a}%
\end{equation}%
\begin{equation}
\sigma=925.5\text{ MeV/fm} \label{ec9b}%
\end{equation}
Let us realize that the value of $\sigma$ stays within the uncertainty
interval of the phenomenological string tension extracted from the $\rho,$
$a_{2},...$Regge trajectory as it should. Regarding $\nu\simeq2r_{s}$ note its
similarity to $R_{2}\simeq2R_{1}$ in the QCDSA. In the same manner $\nu$ can
be interpreted as the onset for string breaking to occur with large
probability. We should realize though that for $r>R_{2}=2.5$ fm $G(r)$ keeps
an almost constant value whereas $F(r)$ varies in a coulombic way.

It is interesting to establish, from the comparison of the spectrum obtained
from $V_{conf}(r),$ Eq.(\ref{ec6}), with data, whether the onset for the
states of confinement in our model may have been experimentally reached or
not. These states may be characterized for having a vanishing probability of
presence for $r<r_{c}$ being $r_{c}$ a distance related to the confinement
scale in QCD. From the quantum number standpoint this means meson states with
orbital angular momentum $L$ greater or equal than a value $L_{c}.$ A
comparison of the light-quark meson spectrum obtained from $V_{conf}(r)$ with
data shows that for $L=4$, for instance, the calculated mass is more than
$100$ MeV above the upper limit of the experimental interval. Although this
difference between calculation and experiment increases when decreasing $L,$
or equivalently decreases when increasing $L,$ the significant discrepancy for
$L=4$ might be suggesting that we are still be far from the pure confinement
region, i. e. $L_{c}>>4.$

\subsection{Phenomenological static potential}

In order to accurately describe the known meson spectra the unquenched
confinement potential, $V_{conf}(r),$ Eq.(\ref{ec6}), has to be complemented.
The natural way to do it is through the incorporation of an effective
\textquotedblleft gluonic\textquotedblright\ coulomb\ interaction so that one
recovers at short distances, when the effect of $q\overline{q}$ pairs is
negligible, the quenched (funnel) form of the potential. For the sake of
simplicity we shall assume the same screening factor used for confinement.
Thus the potential reads%
\begin{equation}
V_{sb}(r)=(\sigma r-\frac{\overline{\lambda}_{q}}{r})(1-e^{-\frac{\nu}{r}})
\label{ec10}%
\end{equation}
where $\overline{\lambda}_{q}$ is the \textquotedblleft
gluonic\textquotedblright\ coulomb strength. The subindex $sb$ indicates that
string breaking has been implemented in both terms of the potential$.$
Certainly $\overline{\lambda}_{q}$ keeps some relation with the
quark-quark-gluon coupling $\alpha_{s}$ since the chromoelectric OGE
contribution should be contained in it. However, corrections to the kinetic
and potential energies could be also taken into account through the effective
value of $\overline{\lambda}_{q}.$ These corrections may include for instance
relativistic terms in the kinetic energy and in the OGE potential,
non-perturbative contributions to the confinement term and to the
quark-quark-gluon coupling, etc. Therefore $\overline{\lambda}_{q}$ has to be
considered as a free parameter. To fix it from data we calculate the high
excited light-quark meson spectrum and require that the known states with a
high orbital angular momentum, $L=4$ for instance$,$ for which we expect the
static approximation works well, are reproduced. The results for
$\overline{\lambda}_{u}=1065$ MeV.fm are presented in Table I. We have used
for the multiplets the quantum numbers notation $(L,n_{r}),$ $n_{r}:$ radial
quantum number, as derived from the solution of the Schr\"{o}dinger equation.
The mass in a multiplet is denoted as $M_{L,n_{r}}.$ Only states giving rise
to $<r^{2}>^{1/2}\gtrsim2.1$ fm, for which the static approximation makes
sense, and for which there are well established experimental candidates, are
considered. \begin{table}[ptb]%
\begin{tabular}
[c]{|c|c|c|c|c|}\hline
($L$, $n_{r}$) & $<r^{2}>^{1/2}$ & $M_{L,n_{r}}$ & ($(M)_{L,n_{r}})_{CBC\text{
}}$ & ($(M)_{L,n_{r}})_{PDG\text{ }}$\\
& fm & MeV & MeV & MeV\\\hline
(5,1) & $3.8$ & $2432$ &  & $a_{6}(2450\pm130)$\\\hline\hline
(4,1) & $2.8$ & $2281^{\dagger}$ & $2262\pm28$ & $\rho_{5}(2330\pm
35)$\\\hline\hline
(3,2) & $3.4$ & $2302$ & $2258\pm38$ & \\\hline\hline
(2,3) & $3.9$ & $2329$ & $2248\pm37$ & $\rho_{3}(2250)$\\\hline\hline
(2,2) & $2.5$ & $2089$ & $1980\pm23$ & $\rho_{3}(1990\pm20)$\\\hline\hline
(1,4) & $4.3$ & $2359$ & $2219\pm43$ & \\\hline
(1,3) & $2.9$ & $2143$ & $1947\pm47$ & \\\hline
\end{tabular}
\caption{Calculated masses $M_{L,n_{r}}$ and rms-radii $<r^{2}>^{1/2}$ for
$(L,n_{r})$ multiplets from $V_{sb}(r)$ with $m_{u}=340$ MeV, $\sigma=925.5$
MeV/fm, $\nu=2.3$ fm and $\overline{\lambda}_{u}=1065$ MeV.fm. Experimental
average masses as in \cite{Gon08} from references \cite{Bug04} ,
($(M)_{L,n_{r}})_{CBC\text{ }},$ and \cite{PDG08}, ($(M)_{L,n_{r}})_{PDG\text{
}}$, are shown for comparison. The superindex $\dagger$ in the (4,1)
calculated mass indicates the average mass value chosen to fix $\overline
{\lambda}_{u}.$}%
\end{table}The ordering of the states has been chosen to make clear the bias
of the results: the lower the $L$ the bigger the difference between calculated
masses and data.

This deficiency can be corrected in an \textit{ad hoc} manner by introducing
an additional $L-$ dependent factor in the \textquotedblleft
gluonic\textquotedblright\ coulomb term so that%
\begin{align}
V(r)  &  =\left(  \sigma r-\frac{\lambda_{q}\left(  1-\sqrt{\frac
{\overrightarrow{L}^{2}}{L_{c}(L_{c}+1)}}\right)  }{r}\right)  (1-e^{-\frac
{\nu}{r}})\text{ \ if }L\leqslant L_{c}\label{ec11}\\
V(r)  &  =V_{conf}(r)\text{ \ \ if \ }L\geq L_{c}\nonumber
\end{align}
This form for $V(r)$ satisfies effectively the requirement that for $L\geq
L_{c}$ the contributions to the energy from other terms in the potential
different than the confining interaction $V_{conf}(r),$ Eq.(\ref{ec6}), are
negligible. Moreover, the potential for states with $L$ close below $L_{c}$
differs little from $V_{conf}(r)$ as it should. Note also that the lower the
$L$ the bigger the probability for short quark-antiquark separations and the
bigger the relativistic corrections to the potential and kinetic energies.
Therefore the value of $\lambda_{q}$ (corresponding to the strength for $L=0$)
may be incorporating, at least to some extent, such corrections in an
effective manner. Actually the same discussion done for $\overline{\lambda
}_{u}$ can be repeated here about the effective character of the
\textquotedblleft gluonic\textquotedblright\ coulomb strength $\lambda_{u}$.
Therefore any attempt to identify our $\lambda_{u}$ with the coefficient of
the chromoelectric coulomb potential obtained from the OGE in QCD is risky.

The calculated masses from $V(r),$ for $\lambda_{u}=1600$ MeV.fm$\ $and
$L_{c}=16,$ and their comparison to data, are shown in Table II where the
states have been now ordered according to their sizes. \begin{table}[ptb]%
\begin{tabular}
[c]{|c|c|c|c|c|}\hline
($L$, $n_{r}$) & $<r^{2}>^{1/2}$ & $M_{L,n_{r}}$ & ($(M)_{L,n_{r}})_{CBC\text{
}}$ & ($(M)_{L,n_{r}})_{PDG\text{ }}$\\
& fm & MeV & MeV & MeV\\\hline
(5,1) & 3.8 & 2432 &  & $2450\pm130$\\
&  &  &  & \\
&  &  &  & a$_{6}(2450)$\\\hline\hline
(1,4) & 3.7 & $2256$ & $2219\pm43$ & \\
&  &  & b$_{1}(2240)$ & \\
&  &  & a$_{1}(2270),$a$_{2}(2175)$ & \\\hline\hline
(2,3) & $3.5$ & $2252$ & $2248\pm37$ & \\
&  &  & $\pi_{2}(2245),\rho(2265)$ & \\
&  &  & $\rho_{2}(2225),\rho_{3}(2260)$ & $\rho_{3}(2250)$\\\hline\hline
(3,2) & $3.1$ & $2250$ & $2258\pm38$ & \\
&  &  & b$_{3}(2245),$a$_{2}(2255)$ & \\
&  &  & a$_{3}(2275),$a$_{4}(2255)$ & \\\hline\hline
(4,1) & $2.7$ & $2254$ & $2262\pm28$ & $2330\pm35$\\
&  &  & $\pi_{4}(2250),\rho_{3}(2260)$ & \\
&  &  & $\rho_{4}(2230),\rho_{5}(2300)$ & $\rho_{5}(2350)$\\\hline\hline
(1,3) & $2.5$ & $1967$ & $1947\pm47$ & \\
&  &  & b$_{1}(1960)$ & \\
&  &  & a$_{1}(1930),$a$_{2}(1950)$ & \\\hline\hline
(2,2) & $2.2$ & $1956$ & $1980\pm23$ & \\
&  &  & $\pi_{2}(2005),\rho(2000)$ & \\
&  &  & $\rho_{2}(1940),\rho_{3}(1982)$ & $\rho_{3}(1990)$\\\hline
\end{tabular}
\caption{Calculated masses and rms-radii from $V(r)$ with $m_{u}=340$ MeV,
$\sigma=925.5$ MeV/fm, $\nu=2.3$ fm, $\lambda_{u}=1600$ MeV.fm and $L_{c}=16$.
Notation as in Table I. The experimental candidates to be members of the
multiplets are also indicated. }%
\end{table}As can be seen the agreement is remarkable. We should not forget
though that the values of the parameters, $\nu,$ $\lambda_{u}\ $and $L_{c}$
have been fixed from the set of data in Table II. Regarding the other
parameters of the model let us recall that the value of $\sigma$ comes from an
external input: the phenomenological analysis of Regge trajectories involving
lowly excited light-quark mesons. As for $m_{u}$ the value chosen corresponds
to the average dynamic mass generated by spontaneous symmetry breaking in the
energy region under consideration, see \cite{Gon08} and references therein.

Let us realize that in most multiplets the difference in mass between members
of the same multiplet is quite small what can be interpreted as the absence of
significant spin-orbit and tensor contributions for the large sized states
considered. On the other hand the calculated meson states become less
relativistic when increasing $(L+n_{r}).$ So $p/m$ goes from $\simeq1.5$ for
$(L+n_{r})=4$ to $\simeq1$ for $(L+n_{r})=6.$ It is then interesting to give
the model predictions for higher $(L,n_{r})$ multiplets for which the
non-relativistic treatment becomes less effective. The average values, from
the calculated masses corresponding to the different $(L,n_{r})$ combinations
giving the same $(L+n_{r}),$ are listed in Table III. \begin{table}[ptb]%
\begin{tabular}
[c]{|c|c|c|}\hline
($L+n_{r}$) & $<r^{2}>^{1/2}$ & $M_{L,n_{r}}$\\
($L,n_{r}$) & fm & MeV\\\hline
6 & $4.6\pm0.8$ & $2421\pm11$\\
(1,5), (2,4), (3,3) &  & \\
(4,2), (5,1) &  & \\\hline
7 &  & \\
(1,6), (2,5), (3,4) & $6.4\pm0.7$ & $2524\pm8$\\
(4,3), (5,2), (6,1) &  & \\\hline
8 &  & \\
(1,7), (2,6), (3,5) & $8.6\pm0.6$ & $2592\pm6$\\
(4,4), (5,3), (6,2), (7,1) &  & \\\hline
9 &  & \\
(1,8), (2,7), (3,6), (4,5) & $11.0\pm0.5$ & $2638\pm5$\\
(5,4), (6,3), (7,2), (8,1) &  & \\\hline
\end{tabular}
\caption{Predicted masses and rms radii from $V(r)$ for some $(L,n_{r})$
multiplets with $L+n_{r}\geq6.$ Parameters as in Table~II.}%
\end{table}A look at the Table shows the quite small difference with the
predictions given in \cite{Gon08} as could be expected from the same
asymptotic behavior of the potentials employed and the large meson radii
involved. It should be also added that the limiting mass for the light-quark
meson spectrum evaluated in \cite{Gon08}\ remains almost unaltered as it is
given by
\begin{equation}
(M_{Limit})_{u\overline{u}}=m_{u}+m_{\overline{u}}+\sigma\nu=2809\text{ MeV}
\label{ec12}%
\end{equation}

\section{Heavy quarkonia}

The proposed form for the screened potential, $V(r),$ should be tested in
other meson sectors. Heavy quarkonia, in particular the non-relativistic
bottomonium, constitutes the ideal laboratory to do it since the static
approximation is expected to be valid for the whole spectrum. To apply $V(r)$
to different meson sectors a criterion to fix the values of the parameters has
to be established. As usual we shall assume that the string tension $\sigma$
is quite approximately flavour independent. Regarding $\nu$ its value has to
do with the screening effect caused dominantly by light sea quark-antiquark
pairs. Consequently it seems\ reasonable to take the same value for it in the
different meson sectors. The universality can be tentatively extended to
$L_{c}$ given its connection to $r_{c},$ or equivalently to the confinement
scale in QCD$.$ Our results will justify this extension. As for the
\textquotedblleft gluonic\textquotedblright\ coulomb strength, $\lambda_{q}$
and the quark mass, $m_{q},$ they will be fixed to get the correct splitting
and masses of two chosen states.

Let us remark that we are dealing with a spin independent potential$.$ For
$s-$ waves we can assume that the experimental energy difference between spin
singlet and spin triplet states (from $V(r)$ they are degenerate) comes mainly
from the spin-spin interaction. Taking into account that the correction for
the spin singlet is in absolute value three times bigger than for the spin
triplet, we shall consider our calculated $s-$ wave states to be approximately
describing spin-triplets. For $p$ and $d-$ waves spin-orbit and tensor
interactions may give significant contributions to the mass. If we remind that
for light-quark mesons this kind of contributions were suppressed for rms
radii, as calculated in our model, greater than 2 fm, we can expect a similar
suppression in heavy quarkonia to take place perhaps at shorter distances
since the strength of the spin-orbit and tensor interactions decreases when
increasing the mass of the quark.

\subsection{Charmonium}

In order to fix $\lambda_{c}$ and $m_{c}$ we should rely on the highest well
established excitations with an unambiguous quantum numbers assignment. We
only have $J/\psi(1s),$ $\psi(2s)$ and $\chi_{c0,c1,c2}(1p).$ As our model
does not contain either spin-orbit or tensor interactions that can give
account of the important mass splitting ($140$ MeV) in the $1p$ multiplet we
should choose $J/\psi(1s)$ and $\psi(2s)$. However, this choice is problematic
since the rms radius for $J/\psi$ obtained from the fixed parameters,
$<r^{2}>^{1/2}=0.4$ fm, would not satisfy the static condition, ($<r^{2}%
>^{1/2})_{c\overline{c}}\gtrsim0.6$ fm, showing that the static approximation
is invalid for $J/\psi$. Instead we shall take for granted the conventional
assignment of $\psi(4040)$ to $\psi(3s)$ and choose $\psi(2s)$ and $\psi(3s)$
as referents to fix the parameters. Notice that attributing the mass
difference $M\left[  \psi(2s)\right]  -M\left[  \eta(2s)\right]  \simeq49$ MeV
to the spin-spin interaction the error in the determination of $M\left[
\psi(2s)\right]  $ due to the non consideration of such interaction is only of
$12$ MeV. For $M\left[  \psi(3s)\right]  $ we expect even a lower error.

The results for the static $c\overline{c}$ spectrum for $\lambda_{c}=157$
MeV.fm and $m_{c}=1448$ MeV are shown in Table IV as compared to data through
a tentative quantum numbers assignment. QCDSA results for the $n_{r}L$ states
are also shown for comparison.

\begin{table}[ptb]%
\begin{tabular}
[c]{ccccc}%
$n_{r}L$ & $\left\langle r^{2}\right\rangle ^{1/2}$ & $M_{L,n_{r}}$ &
($M_{L,n_{r}})_{PDG}$ & ($M_{L,n_{r}})_{SA}$\\\hline
& fm & MeV & MeV & MeV\\\hline
$1s$ &  &  & $3096.916\pm0.011$ & 3105\\
$2s$ & 0.9 & 3686$^{\dagger}$ & $3686.09\pm0.04$ & 3678\\
$1d$ & 1.0 & 3869 & $3772.92\pm0.35$ & 3800\\
$3s$ & 1.4 & 4039$^{\dagger}$ & $4039\pm1$ & 4078\\
$2d$ & 1.5 & 4148 & $4153\pm3$ & 4156\\
$4s$ & 2.0 & 4263 & $4263_{-9}^{+8}$ & 4398\\
$3d$ & 2.1 & 4335 & $4361\pm18^{Be}$ & 4464\\
&  &  & $4324\pm24^{Ba}$ & \\
$5s$ & 2.6 & 4417 & $4421\pm4$ & 4642\\
$4d$ & 2.8 & 4468 &  & 4690\\
$6s$ & 3.3 & 4528 &  & 4804\\
$5d$ & 3.5 & 4565 &  & \\
$7s$ & 4.1 & 4611 &  & \\
$6d$ & 4.4 & 4639 &  & \\
&  &  & $4664\pm16^{Ba}$ & \\
$8s$ & 5.0 & 4674 &  & \\\hline\hline
$1p$ & 0.7 & 3574 & $\chi_{c2}(3556.20\pm0.09)$ & \\
&  &  & $\chi_{c1}(3510.66\pm0.07)$ & \\
&  &  & $\chi_{c0}(3414.75\pm0.31)$ & \\
$2p$ & 1.2 & 3965 & $\chi_{c2}(3929\pm5\pm2)$ & \\
$3p$ & 1.8 & 4212 &  & \\
$4p$ & 2.4 & 4380 &  &
\end{tabular}
\caption{Calculated $c\overline{c}$ masses and rms radii from $V(r).$ The
superindex $\dagger$ indicates the masses used to fix $\lambda_{c}=157$ MeV.fm
and $m_{c}=1448$ MeV. Masses for experimental candidates, $(M_{L,n_{r}}%
)_{PDG},$ have been taken from \cite{PDG08} unless otherwise stated by means
of a superindex: $Be$ for Belle data \cite{Bel07}, $Ba$ for BaBar data
\cite{Ba07}. For $p$ waves we quote the $np_{0}$, $np_{1}$ and $np_{2}$
states. Masses calculated in the QCD string approach \cite{ccBad08},
$(M_{L,n_{r}})_{SA},$ are also shown for comparison. }%
\end{table}As expected the lowest $p$ and $d$ states are not well reproduced.
For $1p,$ with $<r^{2}>^{1/2}=0.7$ fm, the discrepancy goes from 20 MeV for
$\chi_{c2}$ to 160 MeV for $\chi_{c0}.$ For $1d$, with $<r^{2}>^{1/2}=1.0$ fm,
the calculated mass differ about 100 MeV from the only known experimental
candidate. The situation improves extraordinarily for $<r^{2}>^{1/2}%
\gtrsim1.5$ fm since the calculated masses for the $2d,$ $4s,$ $3d$ and $5s$
states can be put in perfect correspondence with experimental candidates (the
resonances $Y(4360)$ from Belle and $Y(4324)$ from BaBar are assumed to
correspond to the same state). This is a very distinctive feature of our model.

It is noteworthy that $X(4260)$ appears as a natural $4s$ state (instead in
the QCDSA the $4s$ state is identified with $\psi(4415))$. Actually the
reluctancy to assign $4s$ quantum numbers to $X(4260)$ comes to some extent
from the much higher mass predicted from conventional charmonium models
\cite{Eic80} since experimental data might be accommodated by making such a
choice \cite{Fel05}. Let us also point out that in our model the $Y(4660)$
reported only by Belle could also correspond to the overlap of the
energetically close $7s$ and $8s$ states. Our $6s$ model state at $4528$ MeV
would be missed as well as other $ns$ states with $n\geq9.$ These excitations
would be very close in energy what could make difficult their experimental
disentanglement despite the fact that the limiting mass of the spectrum is
still quite far above
\begin{equation}
(M_{Limit})_{c\overline{c}}\simeq m_{c}+m_{\overline{c}}+\sigma\nu=5025\text{
MeV} \label{ec13}%
\end{equation}
It should be also remarked the quite relativistic character of the fitted
spectrum with ($p_{c}/m_{c})^{2}\sim0.25-0.16$. Nonetheless the values of the
wave functions at the origin for $ns$ states ($n:2,3,4,5)$ differ at most a
15\% from the ones obtained from the solution of the Salpeter equation in the
QCDSA \cite{ccBad08}. Hence quite similar results (within a 20\% of
difference) would be obtained for the di-electron widths and the same
conclusion inferred: the measured values for $ns$ states ($nd$ ($n:1,2))$ are
systematically smaller (much bigger) than the calculated ones. This can be
explained by the presence of $s-d$ states mixing as a consequence of their
coupling to open channels. This mixing would significantly modify the values
of the $s$ and $d$ wave functions at the origin. On the other hand the very
good fit obtained for the spectrum without mixing suggests that this should
not have any significant effect on the calculated masses of the corresponding
$s$ and $d$ states. Both features can be understood by realizing that
di-electron widths are sensitive to the wave functions at the origin whereas
spectral masses are more related to their long distance behavior.

\subsection{Bottomonium}

A parallel analysis to the one just carried out for charmonium can be done for
bottomonium. As $ns$ states up to $n=4$ have been experimentally identified we
choose $\Upsilon(3s)$ and $\Upsilon(4s)$ to fix the parameters $\lambda_{b}$
and $m_{b}$. From $M\left[  \Upsilon(3s)\right]  =10355$ MeV and $M\left[
\Upsilon(4s)\right]  =10579$ MeV we find $\lambda_{b}=102.6$ MeV.fm and
$m_{b}=4795.5$ MeV. The results for the spectrum are shown in Table V and
assigned to data. For comparison results from the QCDSA are also listed. For
the sake of completeness it should be\ pointed out that the results for $ns$
states with an \textquotedblleft intermediate\textquotedblright\ model based
on the asymptotically constant screened potential of Eq.(\ref{ec4})
\cite{Gon03} lye in between ours and the QCDSA ones. On the other hand quark
potential models not incorporating screening \cite{Eic80} predict much larger
energy splittings for high $n_{r}.$

\begin{table}[ptb]%
\begin{tabular}
[c]{ccccc}%
$n_{r}L$ & $\left\langle r^{2}\right\rangle ^{1/2}$ & $M_{L,n_{r}}$ &
($M_{L,n_{r}})_{PDG}$ & ($M_{L,n_{r}})_{SA}$\\\hline
& fm & MeV & MeV & MeV\\\hline
$1s$ & 0.2 & 9458 & $9460.30\pm0.26$ & 9453\\
$2s$ & 0.5 & 10037 & $10023.26\pm0.31$ & 10010\\
$1d$ & 0.6 & 10218 & $10161.1\pm1.7$ & 10144\\
$3s$ & 0.8 & 10355$^{\dagger}$ & $10355.2\pm0.5$ & 10356\\
$2d$ & 0.8 & 10471 &  & 10446\\
$4s$ & 1.1 & 10579$^{\dagger}$ & $10579.4\pm1.2$ & 10630\\
$5s$ & 1.4 & 10748 &  & 10862\\
$6s$ & 1.7 & 10880 & $10865\pm8$ & 11067\\
&  &  & $10876\pm2^{Ba}$ & \\
$7s$ & 2.0 & 10986 & $10996\pm2^{Ba}$ & 11240\\
&  &  & $11019\pm8$ & \\
$8s$ & 2.4 & 11073 &  & \\
$9s$ & 2.7 & 11144 &  & \\
$10s$ & 3.1 & 11205 &  & \\
$11s$ & 3.6 & 11256 &  & \\\hline\hline
$1p$ & 0.4 & 9970 & $\chi_{b2}(9912.21\pm0.57)$ & 9884\\
&  &  & $\chi_{b1}(9892.78\pm0.57)$ & \\
&  &  & $\chi_{b0}(9859.44\pm0.57)$ & \\
$2p$ & 0.7 & 10300 & $\chi_{b2}(10268.65\pm0.72)$ & 10256\\
&  &  & $\chi_{b1}(10255.46\pm0.72)$ & \\
&  &  & $\chi_{b0}(10232.5\pm0.9)$ & \\
$3p$ & 1.0 & 10535 &  & 10541
\end{tabular}
\caption{Calculated $b\overline{b}$ masses and rms radii from $V(r)$ with
$\lambda_{b}=102.6$ MeV.fm and $m_{b}=4795.5$ MeV. Notation as in Table IV.
The superindex Ba indicates now recent BaBar data \cite{Ba08}. Masses
calculated in the QCD string approach are taken from \cite{bbBad09}.}%
\end{table}Again the $1p,$ $2p$ and $1d$ states are not well described. Now
for $1p$ ($2p)$ with $<r^{2}>^{1/2}=0.4$ fm (0.7 fm), the discrepancy goes
from 60 MeV (30 MeV) for $\chi_{b2}$ to 110 MeV (70 MeV) for $\chi_{b0}.$ For
$1d$, with $<r^{2}>^{1/2}=0.6$ fm, the calculated mass differ about 50 MeV
from the only known experimental candidate. Unfortunately we have not at
disposal data for higher radial $p$ or $d$ excitations to fix a value for the
rms radius beyond which spin dependent contributions are negligible. If we
assume a correct prediction for the $3p$ state this radius would be of $\sim1$ fm.

A very good correspondence between calculated and experimental masses
(difference of 15 MeV at most) is found for $1s,$ $2s,$ $6s$ and $7s$ if the
$\Upsilon(10860)$ is assigned to $\Upsilon(6s)$ (not to $\Upsilon(5s)$ as
usually done) and $\Upsilon(11020)$ to $\Upsilon(7s)$. Notice that the recent
measurements by BaBar give 10876 MeV and 10996 MeV for the masses of these two
resonances. Moreover the $\Upsilon(11020)$ appears in \cite{Lov85} as a peak
between $10990$ MeV and $11060$ MeV what is compatible with being the overlap
of our $7s$ and $8s$ states.

It should be emphasized that the assignment of $\Upsilon(10860)$ to
$\Upsilon(6s)$ implies the existence of a $\Upsilon(5s)$ resonance with a mass%
\begin{equation}
M\left[  \Upsilon(5s)\right]  \sim10748\pm15\text{ MeV} \label{ec14}%
\end{equation}
what can be considered as a main prediction (the quoted error of 15 MeV has
been estimated from Table V) and at the same time as a stringent test of our
potential model. The presence of this resonance might have some relation with
the experimental shoulder present on the tail of $\Upsilon(4s)$ with a mass of
$10684\pm10\pm8$ MeV and a width of $131\pm27\pm23$ MeV in reference
\cite{Bes85} and a mass between $10670$ MeV and $10730$ MeV in \cite{Lov85}
(see Table I of this reference). In the recent study by BaBar \cite{Ba08}
there appears a small bump around $10700$ MeV not identified as a resonance
(see Fig.1 of this reference) that might have to do with the predicted state.
It should be added that the presence of the close $B_{s}\overline{B_{s}}$
threshold at 10732 MeV may complicate the experimental extraction of this
resonance, if existing.

An additional argument in favour of this resonance can be elaborated from the
comparison of the experimental energy differences between contiguous $s$
excitations in charmonium and bottomonium as done in Table VI.

\begin{table}[ptb]%
\begin{tabular}
[c]{ccccc}%
($n_{r}+1)$ &  & ($M_{0,n_{r}+1}-M_{0,n_{r}})$ &  & ($M_{0,n_{r}+1}%
-M_{0,n_{r}})$\\\hline
&  & $c\overline{c}$ &  & $b\overline{b}$\\\hline
$2$ &  & 589 &  & $563$\\
$3$ &  & 353 &  & $332$\\
$4$ &  & 224$^{\ast}$ &  & $224$\\
$5$ &  & 158$^{\ast}$ &  & \\
$6$ &  &  &  &
\end{tabular}
\caption{Experimental mass differences (in MeV) between $(n_{r}+1)s$ and
$n_{r}s$ states in charmonium and bottomonium. The superindex $\ast$ indicates
that the corresponding difference have been calculated assuming that $X(4260)$
and $\psi(4415)$ are the $4s$ and $5s$ states of $c\overline{c}.$}%
\end{table}The assumption that the $4s-3s$ mass differences in bottomonium
(224 MeV) and charmonium have close values as it is the case for $3s-2s$ and
$2s-1s$ requires for $c\overline{c}$ a $4s$ resonance around 4260 MeV such as
our model predicts. Then, assuming that $X(4260)$ is the $4s$ sate, the
extension of the argument to the $5s-4s$ mass differences (158 MeV in
charmonium) implies the existence of $\Upsilon(5s)$ about $10740$ MeV.
Alternatively, as it is the case in the QCDSA, the $X(4260)$ could be not a
$c\overline{c}$ state and the $\Upsilon(10748)$ not exist but in such a case
the energy difference pattern in charmonium and bottomonium would be very
different (the $4s-3s$ mass difference in charmonium would be 382 MeV against
224 MeV in bottomonium). It should be emphasized that this discrepancy in the
interpretation of the experimental states is directly related to the different
manner at which string breaking is implemented in both models. Then the
experimental confirmation (refutation) of our results would serve to establish
the coulombic (non-coulombic) character of the asymptotic quark-antiquark potential.

Concerning other $ns$ states with $n\geq8$ the small separation in energy
between neighbors suggest important overlaps among them and difficulties for a
separated identification. This may explain the non-identification of any clear
signal for a resonance in the region $11000-11200$ MeV recently explored by
BaBar \cite{Ba08}. Let us realize that the limiting mass of the spectrum is
still quite far above
\begin{equation}
(M_{Limit})_{b\overline{b}}\simeq m_{b}+m_{\overline{b}}+\sigma\nu=11720\text{
MeV} \label{ec15}%
\end{equation}

It is also worthwhile to emphasize the non-relativistic character of the
fitted static spectrum in bottomonium since ($p_{b}/m_{b})^{2}\sim0.1-0.06.$
Then we can tentatively identify $\lambda_{b}=4\left(  \alpha_{s}\right)
_{b}/3$ being $\alpha_{s}$ the quark-quark-gluon coupling at the bottomonium
scale. This gives $\left(  \alpha_{s}\right)  _{b}=0.39$ in agreement with the
value derived from QCD in bottomonium for the $1p$ and $2p$ states
\cite{Ynd95}.

As we are dealing with a non-relativistic system we expect that the wave
functions obtained from the Schr\"{o}dinger equation may accurately give
account of other observables. In particular $s-$ wave splittings and leptonic
(di-electron) widths depend directly on the values of the wave functions at
the origin. Thus in first order perturbation theory the splitting energy
between the triplet $\Upsilon(ns)$ and the singlet $\eta_{b}(ns)$ spin states
is given by%
\begin{equation}
M\left[  \Upsilon(ns)\right]  -M\left[  \eta_{b}(ns)\right]  =\frac{4}%
{3}\left(  \alpha_{s}\right)  _{b}\frac{2}{3m_{b}^{2}}\left\vert R_{n_{r}%
s}(0)\right\vert ^{2} \label{ec16}%
\end{equation}
where $R_{n_{r}s}(0)$ stands for the radial wave function at the origin for
$\Upsilon(ns).$ The resulting splitting for $n_{r}=1$ is
\begin{equation}
M\left[  \Upsilon(1s)\right]  -M\left[  \eta_{b}(1s)\right]  =173\text{ MeV}
\label{ec17}%
\end{equation}
in accord with the experimental value
\begin{equation}
M\left[  \Upsilon(1s)\right]  _{ex}-M\left[  \eta_{b}(1s)\right]  _{ex}%
=160\pm40\text{ MeV} \label{ec18}%
\end{equation}
For $n_{r}=2$ the predicted value is
\begin{equation}
M\left[  \Upsilon(2s)\right]  -M\left[  \eta_{b}(2s)\right]  =70\text{ MeV}
\label{ec19}%
\end{equation}

Regarding the leptonic widths $\Gamma_{e^{+}e^{-}}$ for $n_{r}s$ states they
can be evaluated as \cite{Buc81}
\begin{equation}
\Gamma_{e^{+}e^{-}}(n_{r}s)=\Gamma_{e^{+}e^{-}}^{(0)}(n_{r}s)\left[
1-\frac{16\left(  \alpha_{s}\right)  _{b}}{3\pi}+\Delta(n_{r}s)\right]
\label{ec20}%
\end{equation}
The terms with $(-16\left(  \alpha_{s}\right)  _{b}/3\pi)$ and $(\Delta
(n_{r}s))$ give account of the leading order radiative and higher order
radiative + relativistic corrections to%
\begin{equation}
\Gamma_{e^{+}e^{-}}^{(0)}(n_{r}s)\equiv\frac{4e_{b}^{2}\alpha^{2}}{M_{n_{r}%
s}^{2}}\left\vert R_{n_{r}s}(0)\right\vert ^{2} \label{ec21}%
\end{equation}
where $e_{b}=-1/3$ is the quark electric charge, $\alpha=1/137.036$ the fine
structure constant and $M_{n_{r}s}$ the mass of the $n_{r}s$ state for which
we shall use the experimental value. The calculated leptonic widths are shown
in Table VII. Although the correction $\Delta$ depends on the particular
$n_{r}s$ state we shall consider it, for the sake of simplicity, as an
effective constant. We fix its value from $\Gamma_{e^{+}e^{-}}(\Upsilon
(10580))$ since it corresponds to the highest excitation with well identified
quantum numbers ($4s)$ and we expect the non-relativistic and static
approaches to be more accurate for it than for lower excited states. Then we
get $\Delta=0.22$, one third of the value of the first order radiative
correction $16\left(  \alpha_{s}\right)  _{b}/3\pi=0.66.$

\begin{table}[tbh]%
\begin{tabular}
[c]{cccc}%
$n_{r}L$ &  & $\Gamma_{e^{+}e^{-}}$ & $\left(  \Gamma_{e^{+}e^{-}}\right)
_{exp}$\\\hline
$1s$ &  & 1.7 & $1.340\pm0.018$\\
$2s$ &  & 0.61 & $0.612\pm0.011$\\
$3s$ &  & 0.39 & $0.443\pm0.008$\\
$4s$ &  & 0.27$^{\dagger}$ & $0.272\pm0.029$\\
$5s$ &  & 0.21 & $0.20\pm0.05\pm0.10^{\ast}$\\
$6s$ &  & 0.16 & $0.22\pm0.05\pm0.07^{\ast}$\\
$7s$ &  & 0.13 & \\
&  &  & $0.095\pm0.030\pm0.035^{\ast}$\\
$8s$ &  & 0.11 &
\end{tabular}
\caption{Leptonic widths $\Gamma_{e^{+}e^{-}}$ (in keV) for $b\overline{b}$.
The superindex $\dagger$ indicates the value used to fix $\Delta$. Data from
\cite{PDG08} except for $n_{r}s$ with $n_{r}>3$ taken from \cite{Bes85} and
indicated by a superindex $\ast$. The experimental number between the 7s and
8s states indicates that the resonance $\Upsilon$(11020) in \cite{Bes85}, to
which this number is assigned, could be a result of the overlap of the 7s and
8s states.}%
\end{table}It should be pointed out that the measured $\Gamma_{e^{+}e^{-}%
}(10860)$ in \cite{PDG08} might be contaminated by the hidden $\Upsilon
(10748).$ Instead for $n_{r}\geq5$ data from \cite{Bes85} where a resonance
about $1700$ MeV is taken into account are used.

A look at the Table makes clear that a good agreement with data (10\% of
difference at most) may be achieved except for $\Gamma_{e^{+}e^{-}}(1s)$. This
might have to do either with the $\Delta$ dependence on the $n_{r}s$ state or
with a deficient description of the wave function at the origin for
$\Upsilon(1s)$, the more relativistic state for bottomonium with the more
important spin-spin correction. Indeed a 13\% of decrease in the value of
$\left\vert R_{1s}(0)\right\vert $ would fit the central experimental value of
$\Gamma_{e^{+}e^{-}}(1s)$ (notice that the $1s$ spin splitting would be 130
MeV, still within the experimental uncertainty). It should be also kept in
mind that a systematic deviation of the values of the wave functions at the
origin might be hidden through the effective value of $\Delta.$ Actually the
values we get for $R_{1s}(0)$ and $R_{2s}(0)$ are significantly bigger than
the ones obtained in the QCDSA.

Di-electron widths can be also calculated for $nd$ states but no data are
available. Therefore we will only mention that the calculated values in our
model from (see for instance \cite{bbBad09})
\begin{equation}
\Gamma_{e^{+}e^{-}}^{(0)}(nd)=\frac{25e_{b}^{2}\alpha^{2}}{2m_{b}^{4}%
M_{nd}^{2}}\left\vert R_{nd}^{^{\prime\prime}}(0)\right\vert ^{2} \label{ec22}%
\end{equation}
where $R_{nd}^{^{\prime\prime}}(0)$ stands for the second derivative of the
radial wave function at the origin, are four order of magnitudes smaller than
for $n_{r}s$ states.

For the sake of completeness E1 decay widths are also evaluated. By using a
single quark operator approximation the width can be written as \cite{Rob80}
\begin{equation}
\Gamma_{if}^{E1}=\frac{4}{27}e_{b}^{2}\alpha k_{if}^{3}(2J_{f}+1)D_{if}^{2}
\label{ec23}%
\end{equation}
where $k_{if}$ is the photon energy or momentum, $J_{f}$ the total angular
momentum of the final meson and $D_{if}$ the transition matrix element%
\begin{equation}
D_{if}=\int\limits_{0}^{\infty}dru_{i}(r)\frac{3}{k_{if}}\left[  \frac
{k_{if}r}{2}j_{0}\left(  \frac{k_{if}r}{2}\right)  -j_{1}\left(  \frac
{k_{if}r}{2}\right)  \right]  u_{f}(r) \label{ec24}%
\end{equation}
being $u_{i,f}(r)$ the reduced radial wave functions of the initial and final
mesons and $j_{0},$ $j_{1}$ spherical Bessel functions. The results obtained
for $\Upsilon(2s)\rightarrow\gamma\chi_{bJ}(1p)$ and $\Upsilon(3s)\rightarrow
\gamma\chi_{bJ}(2p)$ are compiled in Table VIII. \begin{table}[tbh]%
\begin{tabular}
[c]{ccc}%
\multicolumn{3}{c}{$b\overline{b}$}\\\hline
Transition & $\Gamma_{E1}$ & $\Gamma_{exp}$\\\hline
$\Upsilon(2s)\rightarrow\gamma\chi_{b_{0}}(1P)$ & 1.7 & $1.2\pm0.2$\\
$\Upsilon(2s)\rightarrow\gamma\chi_{b_{1}}(1P)$ & 2.5 & $2.2\pm0.2$\\
$\Upsilon(2s)\rightarrow\gamma\chi_{b_{2}}(1P)$ & 2.6 & $2.3\pm0.3$\\\hline
$\Upsilon(3s)\rightarrow\gamma\chi_{b_{0}}(2P)$ & 1.9 & $1.2\pm0.2$\\
$\Upsilon(3s)\rightarrow\gamma\chi_{b_{1}}(2P)$ & 3.2 & $2.6\pm0.5$\\
$\Upsilon(3s)\rightarrow\gamma\chi_{b_{2}}(2P)$ & 3.5 & $2.7\pm0.6$%
\end{tabular}
\caption{E1 decay widths for $b\overline{b}$ (in keV) as compared to data from
\cite{PDG08}.}%
\end{table}It has to be reminded that the three calculated $\chi_{bJ}(np)$
states are degenerate in our model. Hence the same wave function is employed
for all of them. This can be justified by assuming that the experimental
masses are explained by the effect of spin dependent interactions calculated
in perturbation theory to the first order. The differences in Table VIII among
the three $\Upsilon(2s)\rightarrow\gamma\chi_{bJ}(1p)$ or the three
$\Upsilon(3s)\rightarrow\gamma\chi_{bJ}(2p)$ decays come from the use of the
non-degenerate experimental $\chi_{bJ}$ masses to evaluate $k_{if}$.

A clear bias of the results is observed: they are systematically higher than
data being the discrepancy more pronounced for $\chi_{b0}(np)$ final states.
As $\chi_{b0}(np)$ states are the ones requiring a bigger spin-dependent mass
contribution in our model, the systematics may be suggesting the need of
implementing $\chi_{bJ}(np)$ wave function corrections.

\section{Summary}

To summarize, a universal form for the quark-antiquark static potential,
incorporating the screening of the color charges by sea pairs, has been
proposed within a Non-Relativistic Quark Model framework. This potential, with
a confining long-distance coulombic behavior, reproduces the highly excited
light-quark meson spectrum and provides a successful spectral description of
charmonium and bottomonium suggesting the assignment of $X(4260)$ to the $4s$
state of $c\overline{c}$ and the existence of a non-cataloged $\Upsilon
(10748)$ resonance corresponding to the $5s$ state of $b\overline{b}.$ These
very distinctive predictions of our model come from the way screening have
been implemented in it. Therefore their experimental confirmation or
refutation could allow to establish the coulombic or non-coulombic character
of the long-distance quark-antiquark static potential.

It should be remarked that the only dependence of the potential on the
particular meson sector comes from the value of an effective \textquotedblleft
gluonic\textquotedblright\ strength. As the light-quark mesons and to a lesser
extent the charmonium are clearly relativistic systems one can tentatively
think that some relevant relativistic corrections could be effectively taken
into account through the value of this parameter. The fact that the
\textquotedblleft gluonic\textquotedblright\ coulomb strength gets
systematically a greater value than the gluonic chromoelectric strength in QCD
seems to point out in this direction. For the non-relativistic bottomonium
this \textquotedblleft gluonic\textquotedblright\ strength can be put in
correspondence with the strength of the chromoelectric one gluon exchange
interaction in QCD or, equivalently, with the quark-quark-gluon coupling
$\alpha_{s}$ at the corresponding $Q^{2}$ scale.

In our non-relativistic treatment the quark and antiquark masses are
parameters of the model. Their values should be added to the binding energies
to get the meson masses. A peculiarity of our potential is the absence of any
additive constant to get acceptable values of the constituent quark masses (in
the sense of being able to give account of other observables such as hadronic
magnetic moments) from the fitted meson masses.

All these features make the effective non-relativistic quark model proposed
very useful to identify excited states from existing experimental candidates
and for assigning quantum numbers to them. Furthermore it can be used to
advance predictions on highly excited states in all meson sectors.

This work has been partially funded by the Spanish Ministerio de Ciencia y
Tecnolog\'{\i}a and UE FEDER under Contract No. FPA2007-65748 and by the
spanish Consolider Ingenio 2010 Program CPAN (CSD2007-00042). It is also
partly funded by HadronPhisics2, a FP7-Integrating Activities and
Infrastructure Program of the EU under Grant 227431.


\begin{thebibliography}{99}                                                                                               %


\bibitem {Bug04}D. V. Bugg, Phys. Rep. \textbf{397}, 257 (2004); A. V.
Anisovich, V. V. Anisovich and A. V. Sarantsev, Phys. Rev. D \textbf{62},
051502 (2000).

\bibitem {PDG08}C.~Amsler \textit{et al.} [Particle Data Group],
Phys.\ Lett.\ B \textbf{667}, 1 (2008).

\bibitem {Swa06}E. S. Swanson, Phys. Rep. \textbf{429}, 243 (2006).

\bibitem {Afo07}S. S. Afonin, Mod. Phys. Lett. A \textbf{22}, 1359 (2007);
Int. J. Mod. Phys. A \textbf{23}, 4205 (2008); arXiv:hep-ph/0707.1291.

\bibitem {Gon08}El Houssine Mezoir and P. Gonz\'{a}lez, Phys. Rev. Lett.
\textbf{101}, 232001 (2008).

\bibitem {Bal01}G. S. Bali, Phys. Rep. \textbf{343}, 1 (2001).

\bibitem {Kuz02}A. M. Badalian and D. S. Kuzmenko, Phys. Rev. D \textbf{65},
016004 (2001).

\bibitem {Bor89}K. D. Born \textit{et al. }Phys. Rev. D \textbf{40}, 1653 (1989).

\bibitem {Din95}Y.-B. Ding, K.-T. Chao and D.-H. Qin, Phys. Rev. D
\textbf{51}, 5064 (1995).

\bibitem {Gon03}P. Gonz\'{a}lez, A. Valcarce, H. Garcilazo and J. Vijande,
Phys. Rev. D \textbf{68}, 034007 (2003).

\bibitem {Dun01}A. Duncan, E. Eichten and H. Thacker, Phys. Rev. D
\textbf{63}, 111501 (2001).

\bibitem {Bad02}A. M. Badalian and B. L. G. Bakker, Phys. Rev. D \textbf{66},
034025 (2002).

\bibitem {Sim02}A. M. Badalian, B. L. G. Bakker and Yu. A. Simonov, Phys. Rev.
D \textbf{66}, 034026 (2002).

\bibitem {ccBad08}A. M. Badalian, B. L. G. Bakker and I. V. Danilkin, Phys.
Atom. Nucl. \textbf{72}, 638 (2009), arXiv:0805.2291.

\bibitem {bbBad09}A. M. Badalian, B. L. G. Bakker and I. V. Danilkin, arXiv:0903.3643.

\bibitem {Bel07}X. L. Wang \textit{et al. }(Belle Collaboration), Phys. Rev.
Lett. \textbf{99}, 142002 (2007).

\bibitem {Ba07}B. Aubert \textit{et al. }(BaBar Collaboration), Phys. Rev.
Lett. \textbf{98}, 212001 (2007).

\bibitem {Eic80}E. Eichten, K. Gottfried, T. Kinoshita, K. D. Lane and T. M.
Yan, Phys. Rev. D \textbf{21}, 203 (1980); S. Godfrey and N. Isgur, Phys. Rev.
D \textbf{32}, 189 (1985).

\bibitem {Fel05}F. J. Llanes-Estrada, Phys. Rev. D \textbf{72}, 031503 (2005).

\bibitem {Ba08}B. Aubert \textit{et al. }(BaBar Collaboration), Phys. Rev.
Lett. \textbf{102}, 012001 (2009).

\bibitem {Lov85}D. M. J. Lovelock \textit{et al., }Phys. Rev. Lett.
\textbf{54}, 377 (1985).

\bibitem {Bes85}D. Besson \textit{et al., }Phys. Rev. Lett. \textbf{54}, 381 (1985).

\bibitem {Ynd95}S. Titard and F. J. Yndur\'{a}in\textit{, }Phys. Lett. B
\textbf{351}, 541 (1995); Phys. Rev. D \textbf{51}, 6348 (1995).

\bibitem {Buc81}W. Buchm\"{u}ller and S.-H. H. Tye, Phys. Rev. D \textbf{24},
132 (1981).

\bibitem {Rob80}D. P. Stanley and D. Robson, Phys. Rev. D \textbf{21}, 3180 (1980).
\end{thebibliography}
\end{document}